# The Use of Artificial Intelligence Tools in Assessing Content Validity: A Comparative Study with Human Experts

Hatice Gurdil[1], Hatice Ozlem Anadol[2], and Yesim Beril Soguksu[3]

[1]*Ministry of Education, Ankara, Türkiye;* [2]*Turkish Ministry of National Education, Vali Hilmi Tolun Middle School, Kahramanmaras, Türkiye;* [3]*College of Foreign Languages, Gazi University, Ankara, Türkiye.*

**Abstract**

In this study, it was investigated whether AI evaluators assess the content validity of B1-level English reading comprehension test items in a manner similar to human evaluators. A 25-item multiple-choice test was developed, and these test items were evaluated by four human and four AI evaluators. No statistically significant difference was found between the scores given by human and AI evaluators, with similar evaluation trends observed. The Content Validity Ratio (CVR) and the Item Content Validity Index (I-CVI) were calculated and analyzed using the Wilcoxon Signed-Rank Test, with no statistically significant difference. The findings revealed that in some cases, AI evaluators could replace human evaluators. However, differences in specific items were thought to arise from varying interpretations of the evaluation criteria. Ensuring linguistic clarity and clearly defining criteria could contribute to more consistent evaluations. In this regard, the development of hybrid evaluation systems, in which AI technologies are used alongside human experts, is recommended.

**Keywords:** ChatGPT, Gemini, Chatfuel, Replika, content validity, AI and Human Evaluations, Hybrid Evaluation Systems

**Introduction**

The assessment of individuals' skills, attitudes, and behaviors is essential for both research and real-world applications in the domains of psychology and education (Cohen & Sverdlik, 2010). The development of tests and scales is crucial to achieving accurate and insightful evaluations. Based on an individual's skills, preferences, and psychological traits, these instruments form the basis for assessing cognitive, emotional, and psychomotor skills. Validity and reliability are two key ideas in test development, and they are carefully considered throughout the complex process of developing these assessment tools (Nunnally & Bernstein, 1994).

Assuring content validity is an essential part of test development. How well a test's or scale's items capture the concept they are meant to measure is known as content validity. In other words, the test must cover every facet of the idea without leaving out any crucial details. High content validity is essential for drawing reliable conclusions about individuals based on their test results. For instance, a well-designed reading comprehension test should include items that assess a range of reading skills, such as understanding vocabulary, analyzing texts, and identifying main ideas (Messick, 1989).

Assessing content validity is frequently accomplished through pilot studies, in which a small sample of people take the exam to see its efficacy. Researchers are able to see items that do not perform as predicted or that do not accurately reflect the construct during these pilot studies. The test is significantly enhanced by the insights gained from these studies before being administered to a broader population. However, pilot studies may not always be feasible, particularly in contexts where time or resources are limited. In such cases, the content validity of the test is assessed through expert consultation (Haynes et al.,1995).

The Content Validity Index (CVI) and the Content Validity Ratio (CVR) are two essential instruments in this procedure. These statistical measures are employed to evaluate the validity of an item within a scale or test, ensuring that it accurately represents the construct being measured. Experts evaluate each item to determine whether it is essential for measuring the specific construct, useful but not critical, or unnecessary. Based on these assessments, the CVR is computed, yielding a numerical result that represents the content validity of every item. In contrast, the Content Validity Index (CVI) consolidates individual Content Validity Ratios (CVRs) to offer a comprehensive evaluation of a test's or scale's content validity (Lawshe, 1975).

A significant factor affecting the validity and reliability of CVR and CVI computations is the number of experts engaged in the process. In general, evaluations are more reliable when a greater number of experts are involved. This is because a broader range of professional perspectives can reduce the impact of individual biases and increase the likelihood that the final measures accurately represent the construct being examined. A large and varied panel of experts can offer a more thorough evaluation, guaranteeing that the test items are representative and pertinent (Mertler & Vannatta, 2016).

In practice, however, assembling a sufficiently large group of qualified experts can be challenging. Securing an adequate number of participants is often challenging due to limited resources and the availability of experts. The absence of sufficient expert input can compromise the validity and reliability of content validity assessments, potentially leading to biased or less accurate evaluations. In such cases, ensuring that scales and tests meet the standards for content validity can pose significant challenges for researchers.

Given these challenges, researchers have explored alternative approaches, including artificial intelligence (AI), which offers a scalable and data-driven method for evaluating test items. Leveraging advanced algorithms and large datasets, AI systems can evaluate test items, compare them against predefined standards, and calculate CVR and CVI values using machine learning and natural language processing techniques. While AI cannot fully replicate the nuanced insights of human experts, it can be particularly valuable when expert input is limited or unavailable. By integrating AI into the validation process, researchers can enhance efficiency and accuracy, with further improvements achieved through human expert evaluations.

This study is significant for several reasons. Firstly, by exploring the potential of AI technologies to evaluate the content validity of English language reading comprehension tests, it contributes to advancements in test development. Validity is a critical aspect of test construction, as it ensures that assessments measure the constructs they are intended to evaluate. Additionally, AI techniques offer a novel approach to validating tests, particularly in scenarios where expert resources are limited or unavailable.

Upon reviewing the studies conducted on the potential of artificial intelligence tools in education, Sihite et al. (2023) evaluated ChatGPT's competence in generating reading comprehension questions for academic texts and emphasized the need for further development in this process. Additionally, Morjaria et al. (2023) examined ChatGPT's performance in short-answer assessment questions, highlighting potential risks related to the validity of such assessments. On the other hand, research conducted in the context of scale development has also made significant contributions to understanding the different dimensions of ChatGPT. Nemt-Allah et al. (2023) designed the "ChatGPT Usage Scale" for graduate students, examining dimensions such as academic writing, task support, and dependency. Lee and Park (2023) developed the ChatGPT Literacy Scale to assess its impact on knowledge transfer and creativity. Moreover, the effect of artificial intelligence tools on personalized learning processes has also been noteworthy. Siegle (2023) suggested that these tools could enrich the learning processes of gifted students, while Lin (2024) proposed that ChatGPT could support self-regulated learning among adult learners. In the field of language learning, Kohnke et al. (2023) discussed the opportunities, controversial aspects, and disadvantages of ChatGPT in language teaching and learning, while Saraswati et al. (2023) found that AI Replika could help students practice with native speakers, though it remained limited in understanding cultural meanings. Furthermore, significant findings have emerged in studies comparing artificial intelligence tools with different AI models and human performance. Kim et al. (2024) revealed that, compared to human scores, ChatGPT showed limitations in terms of accuracy and reliability in second-language writing tasks. Hayawi et al. (2023) investigated the capacity of LLMs (BARD, GPT) to be indistinguishable from human writing, while ul haq Akhoon et al. (2024) examined the effectiveness of ChatGPT and Google Gemini in academic content generation and the associated ethical validation issues. Wachinger et al. (2023) found that ChatGPT's performance in qualitative data analysis largely aligned with that of an experienced researcher. Similarly, Gurdil et al. (2023) observed some inconsistencies in item parameters when examining ChatGPT's effectiveness in data generation, but they found results similar to human data. On the other hand, Tenhundfeld (2023) highlighted the need for further development in ChatGPT's human and AI interaction, whereas Onal and Kulavuz-Onal (2023) assessed ChatGPT's high performance in higher education assessment tasks, stating that these tools play a complementary role to human expertise. Upon reviewing the studies, it is evident that the potential of AI-based tools in education has been examined in terms of validity, reliability, and user experience, with comparisons to human performance. However, no research has been found that explores the potential of artificial intelligence tools in determining content validity. Content validity is a critical factor that assesses how accurately a measurement tool captures the entire domain it is intended to measure. Wynd et al.(2023) demonstrated that low inter-expert agreement in the Content Validity Index is crucial for scale revision. This topic is frequently addressed in studies (Ayre & Scally, 2023; Hinkin & Tracey, 2023; Newman et al., 2023), where the importance of innovative approaches to evaluating content validity is emphasized. However, an examination of the literature reveals that no research has specifically explored the potential of artificial intelligence tools in determining content validity. This gap highlights the need for investigating the effectiveness of AI in content validity assessment.

*Purpose of the Study*

This study aims to explore the potential of artificial intelligence (AI) in assessing the content validity of reading skill items from an B1 English exam. Specifically, the study seeks to examine the level of agreement among AI tools and calculate Content Validity Ratios (CVR) and Item Content Validity Indices (I-CVI) based on the ratings provided by four distinct AI tools and four human experts. In this context, the following research questions were formulated:

1. Are there statistically significant differences among the scores given by all evaluators for the items?
2. How consistent are the ratings among all evaluators?
3. Do the CVR values derived from AI evaluators differ significantly from those provided by human evaluators?
4. Do the I-CVI values derived from AI evaluators differ significantly from those provided by human evaluators?

*Significance of Study*

This study explores whether AI can serve as a scalable alternative to expert-based content validity assessments, reducing time and resource constraints. If AI demonstrates effectiveness, it could provide a scalable and cost-efficient alternative, reducing resource demands and enhancing the accessibility of high-quality test validation processes. The study also aims to compare AI-generated content validity evaluations with those made by a human expert. This comparison is essential for evaluating the effectiveness and reliability of AI tools in educational assessment, identifying their strengths and limitations, and informing future test development practices.

Furthermore, the study ensures that its findings have practical implications for language instructors and item writers. Effective AI-based content validity evaluations could lead to more accurate and representative test designs, ultimately benefiting students by providing more reliable measures of their language abilities. Finally, by focusing on a specific application—content validity assessment—this study contributes to the growing body of research on AI applications in education. Successfully integrating AI into this domain could pave the way for further advancements and innovative uses of AI in educational settings, potentially transforming how tests are designed, administered, and validated. Overall, the study's findings could have a substantial impact on test development practices, resource allocation, and the role of artificial intelligence in educational contexts. Additionally, the results may offer valuable insights into the intersection of technology and assessment, guiding future research and practice in the field.

**Method**

*Development and Implementation of the Test*

In this study, a 25-item test consisting of B1-level English reading texts and questions was used for the purposes of the study. In the preparation phase of the test, it was developed by four expert English language instructors. To ensure the validity and reliability of the test, the questions were reviewed by the coordinator of the testing unit, and necessary revisions were made. The test's achievements are compatible with the Common European Framework of Reference for Languages (CEFR). At this point, the respondents are expected to have reading skills corresponding to the B1 level. This means they should be able to understand the general meaning of different types of texts (newspaper articles, essays, stories, etc.), identify the main ideas and important details in the texts, and make inferences about information not explicitly stated. They should also have vocabulary knowledge appropriate to the B1 level and be able to deduce the meanings of words within the text, as well as understand the organization of the text and the connections between paragraphs.

The B1 level was chosen for the study because it represents an intermediate level of English proficiency and appeals to a wide audience. Individuals at this level can understand a variety of texts they may encounter in daily life and can use English for academic or professional purposes. In addition, the B1 level creates a more homogenous group compared to other levels, which ensures that the results of the study are more reliable.

*Evaluators: Human and AI Tools*

After the test was validated, the 25 items it contained were evaluated by 4 human experts and 4 AI tools. The human experts are experts in the field of English Language Teaching, have at least 5 years of experience and are proficient in CEFR assessment processes. The AI tools used in the study are GPT-4o, Gemini 2.0, Replika, Chatfuel, respectively. Developed by OpenAI, ChatGPT-4o is highly successful in analyzing and interpreting complex expressions naturally, thanks to its extensive language processing capacity. ChatGPT-4o, which stands out as a leading model among current artificial intelligence technologies, can be used to analyze texts from different disciplines. While the basic text processing features of GPT-4o can be used for free, some advanced features such as audio and image processing require a subscription (Ofgang, 2023; OpenAI, 2023).

Gemini 2.0 is an artificial intelligence model developed by Google DeepMind, which enhances the evaluation process by leveraging not only text analysis but also visual and contextual information. The model's high performance in text-based tasks has been a significant factor in its selection for this study (DeepMind, 2023). Replika, on the other hand, is a social chatbot that employs advanced natural language processing (NLP) technologies and is built upon the Generative Pretrained Transformer 3 (GPT-3) model. Its technological infrastructure enables it to generate language responses based on a large dataset derived from user conversations, allowing for contextually rich and nuanced interactions (Pentina, Hancock, & Xie, 2023).

In addition to these tools, Chatfuel was also selected for this study. Chatfuel is an AI-based chatbot that uses natural language processing (NLP) techniques to interpret text and generate responses. While it is commonly used in customer service and automated messaging processes, it is also employed to provide feedback and create interactive learning environments. In their study, Ahmed & Hussein (2020) presented a chatbot design that uses the Chatfuel platform to help Kurdish-speaking individuals find answers through online conversations instead of direct contact with human agents. In this study, the use of AI-supported feedback mechanisms in evaluation processes is explored, and for this reason, Chatfuel has been selected.

*Data Analysis*

In this study, human and AI evaluators assessed each test item by assigning scores ranging from 1 to 3 to measure how well the items reflected the intended content. The scoring system was straightforward: '1' meant the item did not represent the content well; '2' indicated the item needed improvement and '3' suggested the item represented the content effectively. Once all items were evaluated, the scores were organized into a table. These ratings were then used to calculate the Content Validity Ratio (CVR) and Item -Level Content Validity Index (I-CVI) and Content Validity Ratio (CVR) for each item. To check whether there were significant differences in the scores given by the evaluators, a Kruskal-Wallis H test was used. This test is a non-parametric alternative to One-Way ANOVA, ideal for analyzing ordinal data that doesn't follow a normal distribution (MacFarland & Yates, 2016). The analysis was conducted using the *'stats'* package in R (R Core Team, 2024).

Each item was reviewed by 8 evaluators -4 humans and 4 AI tools. To assess the level of agreement between human and AI evaluators and among AI tools themselves, the Fleiss Kappa statistic was applied. Fleiss Kappa is a measure of agreement that works well for categorical data when multiple evaluators are involved (Fleiss, 1971; Hallgren, 2012). The calculations were carried out using the *'irr'* package in R (Matthias et al., 2019). Fleiss Kappa values range from -1 to +1, with +1 indicating perfect agreement. In addition, the content validity of the test items was evaluated using two measures: the Content Validity Ratio (CVR) and the Item Content Validity Index (I-CVI). These measures were calculated based on the scores provided by both human and AI evaluators.

***Content Validity Ratio (CVR):*** CVR, introduced by Lawshe (1975), is a widely used method to assess whether test items adequately represent the intended content. Experts categorize each item as 'Essential', 'Useful but not essential' and 'Not necessary'. Items rated as Essential by a certain number of evaluators are kept, while others are discarded (Almanasreh, 2019). The CVR is calculated using the following formula:

$$CV = \frac{n_e - \frac{N}{2}}{\frac{N}{2}} \tag{1}$$

Where '$n_e$' is the number of evaluators who rated the item as "Essential." and 'N' is the total number of evaluators. The CVR score ranges from -1 to +1: '+1' indicates perfect agreement; '-1' indicates complete disagreement and '0' means half of the evaluators rated the item as "Essential." Based on Lawshe's (1975) cut-off points, items with a CVR of at least 0.99 were included in the final version of the test.

***Content Validity Index (I-CVI):*** The Item-Level Content Validity Index (I-CVI) focuses on evaluating the relevance of individual items. Evaluators rate each item on a 4-point scale: 'Not relevant', 'Somewhat relevant', 'Quite relevant' and 'Highly relevant'(Davis, 1992). Lynn (1986) suggested that smaller panels (e.g., 3-5 evaluators) can use either 3- or 5-point scales. The I-CVI is calculated using this formula:

$$I - CVI = \frac{\text{Number of experts who rated the item as relevant}}{\text{Total number of experts}} \quad (2)$$

When there are fewer than 6 evaluators, Lynn (1986) recommended that the I-CVI should be 1 for the item to be considered valid. Similarly, Polit et al. (2007) noted that there should be 100% agreement among evaluators when the panel includes 3 or fewer experts. In this study, only items with an I-CVI of 1 were included in the final form. Items with an I-CVI below 1 were flagged for revision.

**Findings**

The results of the Kruskal-Wallis test and the effect size ($\varepsilon2$) value, conducted to assess whether there is a significant difference between the mean ranks of the scores obtained from eight different evaluators (Human1, Human2, Human3, Human4, GPT-4o, Gemini 2.0, Replika, and Chatfuel) (Appendix 1), are presented in Table 1.

Table 1. Kruskal-Wallis Test Results and Effect Size for Evaluators

| Evaluators | Rank Mean | $\chi^2$ | Df | p-value | Effect Size |
|---|---|---|---|---|---|
| 8 raters* | 13.0 | 10.182 | 7 | 0.1785 | 0.051 |

*Human 1, Human2, Human3, Human4, GPT-4o, Gemini 2.0, Chatfuel and Replika

When examining Table 1, the results of the Kruskal-Wallis test indicate that there is no significant difference between the evaluators ($\chi^2$ = 10.182, df = 7, p > 0.05, $\varepsilon^2$ = 0.051). This result suggests that all evaluators rated the test items in a similar manner. Additionally, the calculated effect size shows that the difference between evaluators is minimal, supporting the conclusion that there is no statistically significant difference between them. The results of the Fleiss Kappa analysis conducted to examine the consistency among all evaluators are presented in Table 2.

Table 2. Results of Consistency Analysis Among Evaluators

| Evaluators | Kappa value | p-value |
|---|---|---|
| 8 raters* | 0.431 | 0 |

*Human 1, Human2, Human3, Human4, GPT-4o, Gemini 2.0, Chatfuel and Replika

As seen in Table 2, A Fleiss Kappa value of 0.431 indicates moderate agreement among evaluators, suggesting some level of consistency, though differences exist in certain items. In other words, there are certain inconsistencies among the evaluators, but overall, they tended to score consistently. Additionally, the most frequently given score (mode) by all evaluators was calculated to be 3. These findings indicate that the evaluators scored consistently, and both human and AI evaluators showed similar performance in the evaluation processes. Based on the scores given by human and AI evaluators, the Content Validity Ratio

(CVR) and Item Content Validity Index (I-CVI) values were calculated, and the obtained values are provided in Table 3.

Table 3. CVR and I-CVI Values for Human and Artificial Intelligence (AI) Scores

|    | CVR_AI | CVR_Human | I-CVI_AI | I-CVI_Human |
|----|--------|-----------|----------|-------------|
| 1  | 0,5    | 0,5       | 0,75     | 0,75        |
| 2  | 0,5    | -1        | 0,75     | 0           |
| 3  | 1      | 1         | 1        | 1           |
| 4  | 1      | 1         | 1        | 1           |
| 5  | 1      | 0,5       | 1        | 0,75        |
| 6  | -0,5   | -0,5      | 0,25     | 0,25        |
| 7  | 1      | 1         | 1        | 1           |
| 8  | 0,5    | 1         | 0,75     | 1           |
| 9  | 0      | -1        | 0,5      | 0           |
| 10 | 1      | 1         | 1        | 1           |
| 11 | 1      | 1         | 1        | 1           |
| 12 | 1      | 1         | 1        | 1           |
| 13 | 0,5    | -0,5      | 0,75     | 0,25        |
| 14 | 1      | 1         | 1        | 1           |
| 15 | 1      | 0         | 1        | 0,5         |
| 16 | 1      | 1         | 1        | 1           |
| 17 | 1      | 1         | 1        | 1           |
| 18 | 1      | 1         | 1        | 1           |
| 19 | 1      | 1         | 1        | 1           |
| 20 | 1      | 1         | 1        | 1           |
| 21 | 1      | 1         | 1        | 1           |
| 22 | 1      | 1         | 1        | 1           |
| 23 | 1      | 1         | 1        | 1           |
| 24 | 1      | 1         | 1        | 1           |
| 25 | 1      | 1         | 1        | 1           |

When examining the CVR and I-CVI values for Human and AI evaluators, it is observed that that human evaluators validated 18 items, whereas AI evaluators validated 19 items (see Table 3). These findings indicate that both evaluator groups demonstrated similar performance in terms of content validity, with the AI evaluators approving one more item at a higher rate. However, both Human and AI evaluators decided not to finalize items 1, 2, 6, 9, and 13. These items were considered inadequate in terms of content validity by both groups and were not approved. The results show that both human and AI evaluators made parallel assessments on these items, demonstrating a similar negative attitude towards them. Similarly, for items 3, 4, 7, 10, 11, 12, 14, 16, 17, 18, 19, 20, 21, 22, 23, 24, and 25, both groups decided to finalize the items. High approval was observed from both human and AI evaluators, and consistent and parallel results were obtained. This further reinforces the notion that AI can make human-like evaluations, yielding results that are consistent with human assessments.

However, item 8 (Appendix 2) reveals a disagreement among evaluators. Human evaluators found this item appropriate in terms of content validity, while AI evaluators preferred revision or removal. This difference may stem from the AI evaluations' inability to fully grasp the context of the text and insufficiently analyze key elements. AI evaluators tend to rely on syntactic and lexical patterns rather than contextual depth, which may explain why they recommended revision for item 8 while human evaluators likely conducted a more in-depth analysis, considering the societal impact and context of the text. Human evaluators considered cultural and societal context, whereas AI evaluators may have overlooked these aspects.

Additionally, for items 5 and 15 (Appendix 2), AI evaluators found both items appropriate in terms of content validity, while human evaluators adopted a more cautious approach, preferring to recommend revision or removal for these items. In the evaluation of item 5, AI evaluators, based on technical criteria, generally found the item sufficient and suggested its finalization, while human evaluators may have taken a more critical approach, assessing the text in terms of context, linguistic sensitivity, and content quality, and opted for a 'revise' or 'remove' decision. This difference may stem from several factors. First, human evaluators may have been more sensitive to the mismatch of the text with its context, while AI evaluators' standard-based evaluation process may overlook such nuances. Human evaluators may also have examined the linguistic quality of the text, the clarity of expressions, and the appropriateness for the target audience in more detail. Furthermore, the alignment of the text with pedagogical objectives and any deficiencies regarding students' learning outcomes may have influenced human evaluators' decisions. On the other hand, AI evaluators may have assessed the text primarily from a technical sufficiency perspective, while human evaluators focused more on content quality, context, and linguistic sensitivity. This discrepancy may be due to AI systems' tendency to overlook certain linguistic and pedagogical features.

Similarly, in the evaluation of item 15, AI evaluators recommended finalizing the question, while human evaluators decided on 'remove' or 'revise'. Several potential reasons for this difference can be considered. Human evaluators may have adopted a more critical approach, particularly regarding the context of the question and its relevance. For example, the statements about the festival's environmental impact and the connection with local producers in this question require a more complex and multifaceted interpretation. AI evaluators, focusing on the technical aspects of the text, concentrated on specific keywords, whereas human evaluators might have addressed the text's integrity and cultural context from a broader perspective. As a result, the criticisms made by human evaluators may have required a more detailed evaluation of the way the question was phrased and the depth of meaning in the answer.

***Differences Between the Calculated CVR and I-CVI for Evaluators' Scores***

The differences between the Content Validity Ratio (CVR) values calculated for the items based on the scores of human and artificial intelligence (AI) evaluators were analyzed using the Wilcoxon Signed-Rank Test. The analysis revealed that the differences between human and AI evaluators were not statistically significant (W = 335.5, p = 0.5708). This result indicates that there was no significant change in the central tendency (median) between the two groups and, therefore, the CVR values are similar.

Although the results obtained from the CVR analysis show the general agreement on the decisions made regarding the items in terms of content validity, it is necessary to examine the suitability of each item in greater depth using the more detailed measurement, the Item Content Validity Index (I-CVI). In this context, the differences between the I-CVI values calculated from the scores of human and AI evaluators were analyzed using the Wilcoxon Signed-Rank Test. The analysis revealed that there was no statistically significant difference between the I-CVI values calculated by human and AI evaluators (W = 335.5, p = 0.5708). In other words, the decisions made by both evaluator groups regarding the suitability of the items were similar. The findings suggest that human and AI evaluators provided consistent results in content validity assessments and that AI systems offer reliability similar to that of human evaluations.

**Conclusion and Discussion**

This study investigated whether human and AI evaluators made similar evaluations in content validity assessments. The results indicated that for the vast majority of items, all evaluators made similar assessments. Furthermore, no evaluator was found to systematically score higher or lower than others. Additionally, the statistical analysis of the differences in the CVR and I-CVI values calculated for the ratings revealed no significant differences, suggesting that AI evaluators provided evaluations that were aligned and similar to those made by human evaluators. The absence of differences between evaluators and the neutrality of the process indicate that AI evaluators could be a reliable and effective tool for content validity analyses.

AI evaluators, based on technical criteria and adherence to standards, are capable of producing results similar to human evaluators in tasks that require objectivity and speed, offering an alternative that could replace human evaluators in such cases. In the future, the potential of AI evaluators to produce objective and accurate results could be leveraged, allowing content validity analyses to be conducted effectively without the need for human intervention.

In this study, overall, there was a high degree of agreement between evaluators in scoring an item, but differences were observed in some items between AI and human evaluators. These differences suggest that there may be different interpretations of evaluation criteria between the two groups. This indicates that certain items may be interpreted more subjectively, or they may require more detailed explanations. Notably, in some items, AI evaluators provided higher levels of approval, while human evaluators adopted a more cautious approach.

AI evaluators adopt a consistent and rule-based approach focused on technical competence but may overlook contextual and linguistic nuances. In contrast, human evaluators assess texts more critically and comprehensively, taking into account societal, cultural, and pedagogical contexts. It becomes clear that AI evaluation processes need improvement in areas such as contextual analysis, linguistic sensitivity, and alignment with pedagogical goals. The capabilities of human evaluators in these areas are particularly crucial when making critical pedagogical and cultural decisions. For example, in some items, AI evaluators, based on technical standards, considered the items sufficient and recommended their inclusion in the final form. On the other hand, human evaluators, taking into account more detailed aspects such as linguistic sensitivity, context, and content quality, made decisions of 'revision' or 'removal.' These differences may arise from the different prioritization of evaluation criteria and the inability of AI systems to fully grasp the subtleties and context of language. Especially in cases where context is critical, human evaluators may have conducted deeper analyses, considering the societal impact of the texts and their alignment with pedagogical objectives. In conclusion, it is suggested that the evaluation processes of AI evaluators should be improved in terms of contextual analysis, linguistic sensitivity, and alignment with pedagogical goals.

These examples highlight that evaluators may adopt different perspectives on content validity, and these differences demonstrate that AI evaluators may not always fully reflect the decision-making processes of human evaluators. In this context, collaboration between AI and human evaluators could contribute to more consistent and reliable results in content validity assessments. While AI evaluators have the advantage of making quick and precise decisions based on linguistic cues, human evaluators may offer the potential for broader critical thinking and consideration of pedagogical concerns in a wider context. This situation suggests that while AI evaluators can perform similarly to human evaluators in certain assessment processes, it may not always be possible for them to fully replace human evaluators in all cases. Similarly, Kim et al. (2023) emphasize that human guidance is critical for AI to produce results with similar accuracy and reliability to humans. Additionally, Fjelland (2020) points out that AI cannot fully reflect human intelligence, while Shane (2019) suggests that AI may outperform human intelligence in certain tasks. In this context, it is important to provide AI evaluators with the necessary training, guidelines, and process improvements to work in harmony with human evaluations.

Taking into account the differences between both approaches in evaluation processes, emphasizing the importance of linguistic clarity and distinguishable options, is believed to contribute to AI evaluators performing similarly to human evaluators. In this regard, defining the criteria more clearly and aligning evaluators' perceptions with these criteria could contribute to making the processes more consistent. The development of hybrid systems where AI-based assessment systems work alongside human evaluations could lead to more reliable and consistent results. Such systems could enable more efficient and accurate content validity analyses by leveraging the strengths of both parties. Furthermore, developing training programs and guidance systems that enhance collaboration between AI and human evaluators could help evaluators achieve more consistent results with a shared understanding. Similar to the results of this study, research conducted by Low et al. (2023), Tenhundfeld (2023), and Vassis et al. (2024) also highlights that AI-based systems do not fully replicate human-like evaluation performance. These studies emphasize that for these systems to align with human evaluations, further development is necessary.

Content validity is critically important, especially in situations where a thorough evaluation of the accuracy and validity of specific test items is required. In the literature, it is generally recommended to consult the opinions of at least 3 to 5 experts in content validity assessments. For example, Lawshe (1975) recommends at least three expert opinions, while Polit et al. (2007) suggest that the decision regarding whether an item should be finalized or not is typically made based on feedback from three experts. However, due to the workload of experts and other practical constraints, it is not always possible to reach these experts and receive feedback (Vassis et al., 2024). In such cases, the involvement of AI evaluators may allow for faster and more efficient content validity assessments. Similar to the studies by Low et al. (2023) and Abu Arqub et al. (2024), this study also shows that AI systems can achieve results similar to human evaluators in content validity reviews, indicating that AI evaluators may replace human intervention in a wider range of contexts in the future.

As a result, it can be said that AI evaluators could be an effective tool in technical analyses such as content validity, offering significant advantages, particularly in situations where the number of experts is high and the expertise requirements are demanding. The ability of AI evaluators to produce results parallel to human evaluations indicates that, in some cases, they could serve as an alternative evaluation method that may replace human intervention. This could allow for the use of AI evaluators' speed and efficiency advantages, especially in studies that require a large number of experts and have time constraints. The data processing speed of AI evaluators and their ability to make consistent decisions based on specific criteria can make the evaluation process more efficient. Future research may make progress in defining processes and evaluation criteria more clearly to make AI and human evaluations more aligned. This could help establish AI evaluators as an effective alternative to humans in content validity assessments. Additionally, conducting similar analyses with larger sample groups and different datasets would allow for testing these findings in a broader context. In conclusion, the strengths of AI evaluators can provide efficiency and effectiveness as a complement to human evaluations. However, it should still be acknowledged that human intervention and guidance are important. Future research is expected to improve the understanding of the balance between AI and human evaluations, making it possible for AI to be applied effectively in a wider range of areas.

**Declaration of Conflicting Interests**

The authors declared no potential conflicts of interest with respect to the research, authorship, and/or publication of this article.

**Funding**

The authors received no financial support for the research, authorship, and/or publication of this article.

**ORCID**

Hatice Gürdil  https:// orcid.org/0000-0002-0079-3202

Hatice Özlem Anadol  https:// orcid.org/0000-0002-2881-9806

Yeşim Beril Soğuksu  https://orcid.org/0009-0004-0870-4974

# Appendix

*Appendix 1.* Scores Obtained from Human and AI Evaluators

| Gemini | GPT | Replika | Chatfuel | Human1 | Human2 | Human3 | Human4 |
| --- | --- | --- | --- | --- | --- | --- | --- |
| 3 | 2 | 3 | 3 | 3 | 2 | 3 | 3 |
| 2 | 3 | 3 | 3 | 2 | 2 | 2 | 2 |
| 3 | 3 | 3 | 3 | 3 | 3 | 3 | 3 |
| 3 | 3 | 3 | 3 | 3 | 3 | 3 | 3 |
| 3 | 3 | 3 | 3 | 3 | 2 | 3 | 3 |
| 2 | 3 | 2 | 2 | 2 | 2 | 2 | 3 |
| 3 | 3 | 3 | 3 | 3 | 3 | 3 | 3 |
| 3 | 3 | 2 | 3 | 3 | 3 | 3 | 3 |
| 2 | 3 | 2 | 3 | 2 | 2 | 2 | 2 |
| 3 | 3 | 3 | 3 | 3 | 3 | 3 | 3 |
| 3 | 3 | 3 | 3 | 3 | 3 | 3 | 3 |
| 3 | 3 | 3 | 3 | 3 | 3 | 3 | 3 |
| 3 | 3 | 2 | 3 | 2 | 2 | 2 | 3 |
| 3 | 3 | 3 | 3 | 3 | 3 | 3 | 3 |
| 3 | 3 | 3 | 3 | 2 | 2 | 3 | 3 |
| 3 | 3 | 3 | 3 | 3 | 3 | 3 | 3 |
| 3 | 3 | 3 | 3 | 3 | 3 | 3 | 3 |
| 3 | 3 | 3 | 3 | 3 | 3 | 3 | 3 |
| 3 | 3 | 3 | 3 | 3 | 3 | 3 | 3 |
| 3 | 3 | 3 | 3 | 3 | 3 | 3 | 3 |
| 3 | 3 | 3 | 3 | 3 | 3 | 3 | 3 |
| 3 | 3 | 3 | 3 | 3 | 3 | 3 | 3 |
| 3 | 3 | 3 | 3 | 3 | 3 | 3 | 3 |
| 3 | 3 | 3 | 3 | 3 | 3 | 3 | 3 |

*Appendix 2:* Sample Items from the Test

PART I

**Benefit for customers**

The advantage of having a multicultural workforce is that it can teach people to look at things from a different point of view and solve problems more effectively. Employees have to learn to see things from a different point of view. As a result, this can often lead to a satisfactory solution that they were unable to see before. In addition, the different skills of cultural understanding and languages enable a company to give better customer service around the world.

**5. How does a culturally diverse workforce benefit customers according to Paragraph 4?**
**a)** By enhancing customer support services
**b)** By making innovative products
**c)** By maintaining a good reputation
**d)** By managing cultural differences
**e)** By offering problem solving skills for customers

PART II

Similarly, Oprah Winfrey, a favorite media entrepreneur, is well-known for her extensive network of charitable efforts. From funding poor children's education to supporting women's empowerment programs, Winfrey's behavior exemplifies the power of using one's platform for the wellbeing of positive change.

**8. As stated in Paragraph 2, Oprah Winfrey supports _____ .**

**a)** learning organizations

**b)** children's psychological platforms

**c)** men's empowerment programs

**d)** women's employment training

**e)** public health education

**PART III**

Despite these recommendations, some obstacles such as moldy or stale food may arise. However, adopting some solutions such as freezing or preserving can help extend the shelf life of fresh items, minimising waste. Food festivals can continue to be a source of joy and inspiration while protecting the planet by considering the cost of food waste and embracing environmentally friendly approaches. They give people a chance to try delicious foods, learn about local traditions, and meet other food lovers.

**15. According to Paragraph 4, food festivals _____ .**

**a)** affect the people's overall health

**b)** decline the number of imported foods

**c)** help people meet local food producers

**d)** make people prepare meals cooked in other countries

**e)** reduce the financial impact of wasteful consumption